\documentclass[10pt,aps,twocolumn,prd,notitlepage,amssymb,amsmath,floatfix,nofootinbib,superscriptaddress]{revtex4-1}
\usepackage{epsfig}
\usepackage{bm}
\usepackage{amssymb}
\usepackage{amsmath}
\usepackage{color}
\usepackage{xcolor}
\usepackage{slashed}
\usepackage[colorlinks,linkcolor=blue,anchorcolor=black,citecolor=blue]{hyperref}
\usepackage{multirow}
\usepackage{ulem}
\usepackage{comment}

\allowdisplaybreaks[4]

\begin{document}

\title{Polarized vector meson production in semi-inclusive DIS}

\author{Xue-shuai Jiao}
\affiliation{School of Science, Shandong Jianzhu University, Jinan, Shandong 250101, China}

\author{Kai-bao Chen}
\email{chenkaibao19@sdjzu.edu.cn}
\affiliation{School of Science, Shandong Jianzhu University, Jinan, Shandong 250101, China}

\begin{abstract}
We make a systematic calculation for polarized vector meson production in semi-inclusive lepton-nucleon deep inelastic scattering $e^-N\to e^-VX$.
We consider the general case of neutral current electroweak interactions at high energies which give rise to parity-violating effects.
We present a general kinematic analysis for the process and show that the cross sections are expressed by 81 structure functions.
We further give a parton model calculation for the process and show the results for the structure functions in terms of the transverse momentum dependent parton distribution functions and fragmentation functions at the leading order and leading twist of perturbative quantum chromodynamics.
The results show that there are 27 nonzero structure functions at this order, among which 15 are related to the tensor polarization of the vector meson.
Thirteen structure functions are generated by parity-violating effects.
We also present the result and a rough numerical estimate for the spin alignment of the vector meson.
\end{abstract}

\maketitle

\section{Introduction}
Parton distribution functions (PDFs) and fragmentation functions (FFs) are related to the hadron structure and hadronization mechanism, respectively.
They are two important quantities in describing high energy reactions (see e.g.,~\cite{Barone:2001sp,DAlesio:2007bjf,Barone:2010zz,Peng:2014hta,Chen:2015tca,Metz:2016swz,Anselmino:2020vlp} for recent reviews).
In semi-inclusive reaction processes, e.g., lepton-nucleon semi-inclusive deep inelastic scattering (SIDIS) with hadron production at small transverse momentum, the transverse momentum dependent (TMD) factorization applies~\cite{Collins:1981uk,Collins:1981va,Collins:1984kg,Ji:2004wu,Ji:2004xq}.
The sensitive observables in experiments are often different azimuthal asymmetries that are theoretically expressed by the convolutions of TMD PDFs and FFs in general.
Therefore, processes such as SIDIS, Drell-Yan, or semi-inclusive hadron production in electron-positron annihilation give us opportunities for accessing the three-dimensional hadron structure and hadronization mechanism~\cite{Levelt:1994np,Levelt:1993ac,Kotzinian:1994dv,Tangerman:1994eh,Mulders:1995dh,Boer:1997mf,Kotzinian:1997wt,Boer:1997nt,Boer:1997qn,Bacchetta:2000jk,Boer:1999uu,Bacchetta:2004zf,Bacchetta:2006tn,Boer:2008fr,Pitonyak:2013dsu,Yang:2016qsf,Liang:2006wp,Liang:2008vz,Gao:2010mj, Song:2010pf, Song:2013sja,Wei:2013csa,Wei:2014pma,Chen:2016moq,Wei:2016far,Yang:2017sxz,Chen:2020ugq}.

One can access the spin dependent TMD FFs in semi-inclusive processes by measuring the polarizations or spin dependent azimuthal asymmetries of the produced hadron.
For example, for $\Lambda$ hyperon production, the polarizations of $\Lambda$ can be measured by its self-analyzing weak decay.
Very similar to hyperon polarizations, the polarizations of vector mesons can also be determined by the angular distribution of their decay products through the strong decay into two spin zero hadrons.
Since the vector mesons are spin-$1$, there will be both vector and tensor polarizations for them.
Measurements have been carried out, e.g., for the $e^+e^-$ annihilation process at the Large Electron-Positron Collider more than two decades ago for the spin alignment of vector mesons~\cite{OPAL:1997vmw,OPAL:1997nwj,DELPHI:1997ruo}, which has attracted much attention (see, e.g.,~\cite{Chen:2016iey} for a recent phenomenological study and references therein).
One can also study hadron polarizations the in SIDIS process; e.g., the $\Lambda$ hyperon polarizations can be studied in $e^-N\to e^-\Lambda X$.
In the QCD parton model and TMD factorization, the $\Lambda$ polarizations in SIDIS are expressed by the convolution of the TMD PDFs with the corresponding spin dependent TMD FFs (see~\cite{Chen:2021zrr} for a recent phenomenological study).
When we consider the final state hadron to be a vector meson $V$ with spin-1, i.e., $e^-N\to e^-VX$, we can access not only the vector polarization dependent FFs but also the tensor polarization dependent FFs~\cite{Chen:2016moq}.

The Electron-Ion Collider (EIC) has been proposed to be built as the next generation collider on deep inelastic scattering with high energy and high luminosity~\cite{Accardi:2012qut,AbdulKhalek:2021gbh}.
It gives us new opportunities for exploring physics on quantum chromodynamics (QCD) and nucleon structure.
Since the beam energy of the EIC is relatively high, it has a wider kinematic coverage.
The momentum transfer $Q^2$ between the incident lepton and the nucleon has a chance to be comparable with $M_Z^2$, i.e., the mass square of the $Z$ boson.
Therefore, parity-violating effects can arise through the interference of the electromagnetic (EM) and weak interactions~\cite{Cahn:1977uu}.
It will also generate new structure functions that are complementary to those from the pure EM interaction.
Experimentally, parity-violating asymmetries in deep inelastic scattering (DIS) experiments were first observed
at SLAC~\cite{Prescott:1978tm,Prescott:1979dh} and have been measured widely~\cite{Aniol:2004hp,Aniol:2005zf,Aniol:2005zg,Armstrong:2005hs,Androic:2009aa, Wang:2013kkc,Wang:2014guo,Spayde:1999qg,Ito:2003mr,Maas:2004dh,Maas:2004ta}.
Proposals for precise measurements are also available~\cite{PVDIS:Jlab6,PVDIS:JLab12,Zhao:2017xej}.
On the theoretical side, electroweak inclusive and semi-inclusive DIS processes have been studied extensively \cite{Cahn:1977uu,Anselmino:1993tc,Ji:1993ey,Anselmino:1994gn,Boer:1999uu,Anselmino:2001ey,deFlorian:2012wk,Moreno:2014kia,Chen:2020ugq}.
However, systematic studies are still lacking.
These include a full kinematic analysis and QCD parton model calculations for the cross section, and the systematic treatment for the hadron polarization effects.

The rest of the paper is organized as follows. 
In Sec.~\ref{sec:Kin}, we present the general form of the cross section for $e^-N\to e^-VX$ in terms of structure functions by carrying out a kinematic analysis.
In Sec.~\ref{sec:HT}, we calculate the hadronic tensor in the QCD parton model and give the results expressed by the convolution of the TMD PDFs and FFs.
In Sec.~\ref{sec:Results}, we give the results for the structure functions as well as the spin alignment of the vector meson in terms of TMD PDFs and FFs.
Section~\ref{sec:Sum} is a summary.

\section{The general form of the cross section and the structure functions}
\label{sec:Kin}

\subsection{The process and notations}
\begin{figure}[htb]
\includegraphics[width=0.45\textwidth]{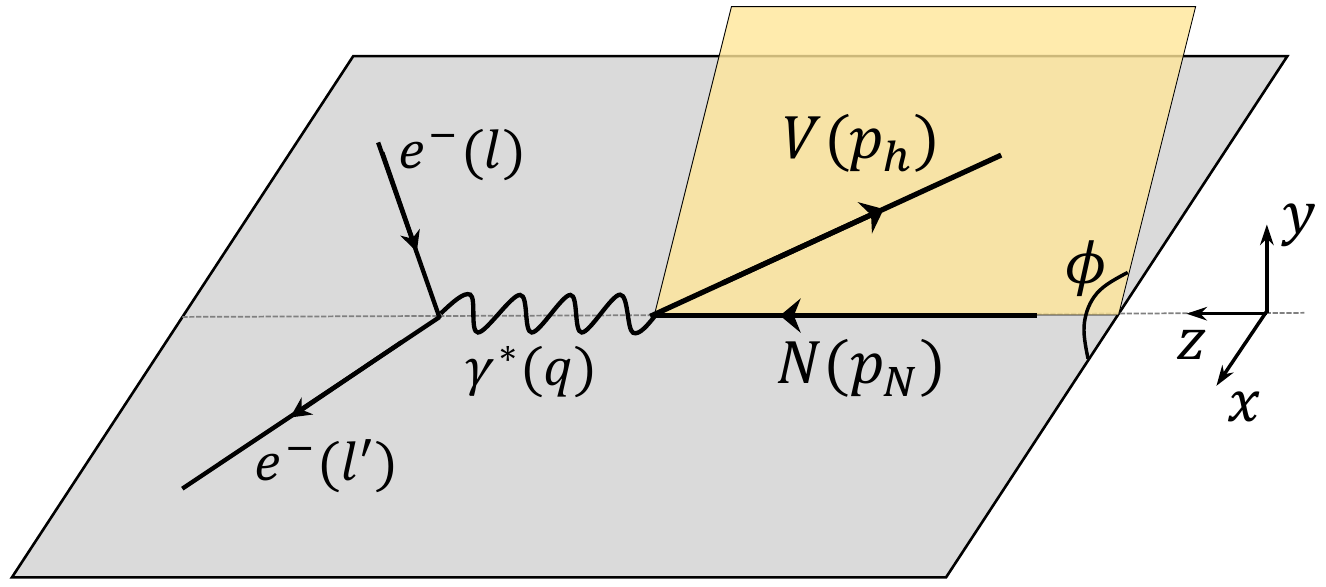}
\caption{The coordinate system for $e^-N\to e^-V X$. The symbols in the brackets denote the four momenta of the particles.}
\label{fig:sidis-kinematics}
\end{figure}

We consider the SIDIS process at high energies with unpolarized electron and nucleon beams, i.e.,
\begin{align}
e^-(l) + N(p_N) \rightarrow e^-(l^\prime) + V(p_h,S) + X,
\end{align}
where $V$ is a vector meson with spin-1.
The momenta of the incident and outgoing particles are shown in the brackets. $S$ denotes the polarizations of the vector meson.
The coordinate system for $e^-N\to e^- VX$ in the photon-nucleon collinear frame is shown in Fig.~\ref{fig:sidis-kinematics}, where the incoming proton and the virtual photon move along the $\pm z$ axis and the $x$ axis is determined by the transverse momenta of the leptons. The azimuthal angle $\phi$ is spanned by the transverse momentum of the vector meson with respect to the transverse momentum of the outgoing electron.

The standard variables for SIDIS are defined as $Q^2 = -q^2$, $x=\frac{Q^2}{2 p_N\cdot q}$, $y=\frac{p_N\cdot q}{p_N \cdot l}$, $z_h = \frac{p_N \cdot p_h}{p_N\cdot q}$, and $s=(p_N+l)^2$.
We will use the light-cone coordinate system, in which a four vector $a^\mu$ is expressed as $a^\mu = (a^+,a^-,\vec a_\perp)$ with $a^{\pm} = (a^0\pm a^3)/\sqrt{2}$ and $\vec a_\perp = (a^1,a^2)$.
The momenta of the particles take the following forms:
\begin{align}
& p_N^\mu = \left(p_N^+,0,\vec 0_\perp \right),\\
& l^\mu = \left( \frac{1-y}{y}xp_N^+, \frac{Q^2}{2xyp_N^+}, \frac{Q\sqrt{1-y}}{y},0 \right),\\
& q^\mu = \left( -xp_N^+, \frac{Q^2}{2xp_N^+}, \vec 0_\perp \right),\\
& p_{h}^\mu = \left(\frac{xp_N^+ \vec p_{h\perp}^2}{z_h Q^2}, \frac{z_h Q^2}{2xp_N^+}, \vec p_{h\perp} \right),\\
& \vec p_{h\perp} = |p_{h\perp}|(\cos\phi, \sin\phi).
\end{align}
We also define two unit light-cone vectors $n^\mu = (0,1,\vec 0_\perp)$ and $\bar n^\mu = (1,0,\vec 0_\perp)$.
With these notations, the transverse metric is given by $g_\perp^{\mu\nu} = g^{\mu\nu} - \bar n^\mu n^\nu - \bar n^\nu n^\mu$.
We will also use the transverse antisymmetric tensor defined as $\varepsilon_\perp^{\mu\nu} = \varepsilon^{\mu\nu\alpha\beta} \bar n_\alpha n_\beta$ with $\varepsilon_{\perp}^{12} = 1$.

We consider the leading order approximation for QED with the exchange of a single virtual photon $\gamma^*$ or a $Z$ boson with momentum $q=l-l^\prime$ between the electron and the nucleon.
The differential cross section is given by
\begin{align}
d\sigma = \frac{2\alpha_{\rm em}^2}{sQ^4} A_{r} 
L_r^{\mu\nu}(l, l^\prime)
W_{r,\mu\nu}(q,p_N,p_h,S)
\frac{d^3 l^\prime d^3 p_h}{2E_{l^\prime} 2E_h}.
\label{eq:Xsec0}
\end{align}
The symbol $r$ can be $\gamma\gamma$, $ZZ$, and $\gamma Z$.
They correspond to the electromagnetic, the weak, and the interference contributions, respectively.
A summation over $r$ is implicit in Eq.~(\ref{eq:Xsec0}), i.e.,
\begin{align}
d\sigma = d\sigma_{\gamma\gamma} + d\sigma_{ZZ} + d\sigma_{\gamma Z}.
\label{eq:TotalXsec}
\end{align}
The $A_r$ factors and the leptonic tensors are determined in perturbative theory.
The $A_r$ factors can be calculated as
\begin{align}
& A_{\gamma\gamma} = e_q^2,\\
& A_{ZZ} = \frac{Q^4}{\left[(Q^2+M_Z^2)^2 + \Gamma_Z^2 M_Z^2 \right] \sin^4 2\theta_W} \equiv \chi,\\
& A_{\gamma Z} = \frac{2e_q Q^2 (Q^2+M_Z^2)}{\left[(Q^2+M_Z^2)^2 + \Gamma_Z^2 M_Z^2 \right] \sin^2 2\theta_W} \equiv \chi_{\rm int},
\end{align}
where $\Gamma_Z$ is the width of the $Z$ boson and $\theta_W$ is the Weinberg angle.
The leptonic tensors are given by
\begin{align}
& L_{\gamma\gamma}^{\mu\nu}(l, l^\prime) 
= 2\left( l^\mu l^{\prime\nu} + l^\nu l^{\prime\mu} - g^{\mu\nu} l\cdot l^\prime \right), \\
& L_{ZZ}^{\mu\nu}(l, l^\prime)
= c_1^e L_{\gamma\gamma}^{\mu\nu}(l, l^\prime) - 2ic_3^e\varepsilon^{\mu\nu l l^\prime}, \\
& L_{\gamma Z}^{\mu\nu}(l, l^\prime)
= c_V^eL_{\gamma\gamma}^{\mu\nu}(l, l^\prime) - 2ic_A^e\varepsilon^{\mu\nu l l^\prime},
\end{align}
where $c_V^e$ and $c_A^e$ are from the weak interaction vertex defined by $\Gamma_\mu^e = \gamma_\mu(c_V^e - c_A^e \gamma_5)$, $c_1^e = (c_V^e)^2 + (c_A^e)^2$, and $c_3^e = 2c_V^e c_A^e$.
We see that $L_{\gamma\gamma}^{\mu\nu}$ and $L_{\gamma Z}^{\mu\nu}$ can be obtained from $L_{ZZ}^{\mu\nu}$ by the replacements of $(c_1^e,c_3^e)\to(1,0)$ and $(c_1^e,c_3^e)\to(c_V^e,c_A^e)$, respectively.
The corresponding hadronic tensors are defined as
\begin{align}
& W_{\gamma\gamma}^{\mu\nu}(q,p_N,p_h,S) = \sum_X \delta^4(p_N + q - p_h - p_X) \nonumber\\
&\qquad\times \langle p_N | J_{\gamma\gamma}^{\mu}(0) |p_h,S;X\rangle \langle p_h,S;X | J_{\gamma\gamma}^{\nu}(0) | p_N \rangle,\\
& W_{ZZ}^{\mu\nu}(q,p_N,p_h,S) = \sum_X \delta^4(p_N + q - p_h - p_X) \nonumber\\
&\qquad \times \langle p_N | J_{ZZ}^{\mu}(0) |p_h,S;X\rangle \langle p_h,S;X | J_{ZZ}^{\nu}(0) | p_N \rangle,\\
& W_{\gamma Z}^{\mu\nu}(q,p_N,p_h,S) = \sum_X \delta^4(p_N + q - p_h - p_X) \nonumber\\
&\qquad \times \langle p_N | J_{ZZ}^{\mu}(0) |p_h,S;X\rangle \langle p_h,S;X | J_{\gamma\gamma}^{\nu}(0) | p_N \rangle, 
\end{align}
where the current operators are given by $J_{\gamma\gamma}^{\mu}(0) = \bar\psi(0) \gamma^\mu \psi(0)$ and $J_{ZZ}^{\mu}(0) = \bar\psi(0) \Gamma_q^\mu \psi(0)$ with $\Gamma_\mu^q = \gamma_\mu(c_V^q - c_A^q \gamma_5)$.

For the phase space factors, in terms of the standard SIDIS variables, we have
\begin{align}
\frac{d^3 l^\prime}{2E_{l^\prime}} = \frac{y s}{4} dx d y d\psi,\qquad \frac{d^3p_h}{2E_h} = \frac{dz_h}{2z_h} d^2 p_{h\perp},
\end{align}
where $\psi$ is the azimuthal angle of $\vec l^\prime$ around $\vec l$.
Therefore the cross section in Eq. (\ref{eq:Xsec0}) can be written as
\begin{align}
\frac{d\sigma}{dx dy dz_h d\psi d^2 p_{h\perp}} = \frac{y \alpha_{\rm em}^2}{4 z_h Q^4} A_{r}
L_r^{\mu\nu}(l, l^\prime) W_{r,\mu\nu}(q,p_N,p_h,S).
\label{eq:Xsec}
\end{align}

\subsection{Cross section in terms of structure functions}
In this subsection, we make a kinematic analysis for the process. 
There is no restriction from parity since we consider both the EM and the weak interaction contributions.
The most general form of the cross section is obtained by adding the contributions from the EM, the weak and the interference terms together.
Formally, the pure weak interaction contribution will generate all the possible structure functions.
The EM and the interference contributions will not introduce new types of structure functions, so we first concentrate on the analysis of the pure weak interaction contribution.

We give the decomposition of the hadronic tensors in terms of basic Lorentz tensors (BLTs).
The general form of the hadronic tensor satisfies the constraints of Hermiticity and current conservation.
The hadronic tensor is divided into a symmetric part and an antisymmetric part, i.e.,
$W_{ZZ}^{\mu\nu} = W_{ZZ}^{S\mu\nu} + W_{ZZ}^{A\mu\nu}$.
More explicitly, we have
\begin{align}
W_{ZZ}^{S\mu\nu} &=\sum_{\sigma,i} W_{\sigma i}^S h_{\sigma i}^{S\mu\nu} + \sum_{\sigma, i} \tilde W_{\sigma i}^S \tilde h_{\sigma i}^{S\mu\nu},\label{eq:WSuv}\\
W_{ZZ}^{A\mu\nu} &=\sum_{\sigma,i} W_{\sigma i}^A h_{\sigma i}^{A\mu\nu} + \sum_{\sigma, i} \tilde W_{\sigma i}^A \tilde h_{\sigma i}^{A\mu\nu},\label{eq:WAuv}
\end{align}
where $h^{\mu\nu}_{\sigma i}$'s and $\tilde h^{\mu\nu}_{\sigma i}$'s represent the space reflection even and odd BLTs, respectively.
They are constructed from available kinematical variables in the process, e.g., $p_N$, $p_h$, $q$, and the polarization vector or tensor.
$W_{\sigma i}$'s are scalar coefficients.
The subscript $\sigma$ specifies the polarizations of the vector meson.

There are nine BLTs for the unpolarized part~\cite{Chen:2016moq}, i.e.,
\begin{align}
h^{S\mu\nu}_{Ui}&=\Big\{g^{\mu\nu}-\frac{q^\mu q^\nu}{q^2}, ~p_{Nq}^\mu  p_{Nq}^\nu,~ p_{hq}^{\mu}  p_{hq}^{\nu},  ~ p_{Nq}^{\{\mu} p_{hq}^{\nu\}}\Big\},\label{eq:hsU} \\
\tilde h^{S\mu\nu}_{Ui}&=\Big\{\varepsilon^{\{\mu q p_N p_h} p_{Nq}^{\nu\}},~ \varepsilon^{\{\mu q p_N p_h} p_{hq}^{\nu\}}\Big\}, \label{eq:thsU} \\
h^{A\mu\nu}_{U}&=\Big\{p_{Nq}^{[\mu} p_{hq}^{\nu]}\Big\},\label{eq:haU} \\
\tilde h^{A\mu\nu}_{Ui}&=\Big\{\varepsilon ^{\mu\nu qp_N},~ \varepsilon ^{\mu\nu qp_h}\Big\}.\label{eq:thaU}
\end{align}
The subscript $U$ denotes the unpolarized part. 
We have defined the four vectors such as $p_{Nq}^\mu \equiv p_N^\mu - q^\mu(p_N\cdot q)/q^2$ satisfying $q\cdot p_{Nq}$ = 0, and similar for $p_{hq}^\mu$.
We have also used the notations $A^{\{\mu}B^{\nu\}} \equiv A^\mu B^\nu +A^\nu B^\mu$, and $A^{[\mu}B^{\nu]} \equiv A^\mu B^\nu -A^\nu B^\mu$ for simplicity.

The vector polarization of the vector meson, similar to spin-$1/2$ hadrons, is described by the helicity $\lambda$ and the transverse polarization vector $S_T^\mu$.
It has been shown in~\cite{Chen:2016moq} that the polarization dependent BLTs can be constructed from the unpolarized BLTs by multiplying the corresponding polarization dependent scalars or pseudoscalars.
Therefore, we have
\begin{align}
h^{S\mu\nu}_{Vi} &=\Big\{[\lambda, (p_{h\perp} \cdot S_T)]\tilde h^{S\mu\nu}_{Ui}, ~\varepsilon_\perp^{p_{h\perp} S_T} h^{S\mu\nu}_{Uj}\Big\}, \label{eq:hsV}\\
\tilde h^{S\mu\nu}_{Vi} &=\Big\{[\lambda, (p_{h\perp} \cdot S_T)]h^{S\mu\nu}_{Ui}, ~ \varepsilon_\perp^{p_{h\perp} S_T} \tilde h^{S\mu\nu}_{Uj} \Big\}, \label{eq:thsV}\\
h^{A\mu\nu}_{Vi} &=\Big\{[\lambda, (p_{h\perp} \cdot S_T)]\tilde h^{A\mu\nu}_{Ui},~ \varepsilon_\perp^{p_{h\perp} S_T}  h^{A\mu\nu}_{U}\Big\},\label{eq:haV}\\
\tilde h^{A\mu\nu}_{Vi} &=\Big\{[\lambda, (p_{h\perp} \cdot S_T)]h^{A\mu\nu}_{U},~ \varepsilon_\perp^{p_{h\perp} S_T}  \tilde h^{A\mu\nu}_{Uj}\Big\}.\label{eq:thaV}
\end{align}
There are in total 27 BLTs for the vector polarization dependent part.

The tensor polarization of the vector meson is described by five independent components.
They are given by a Lorentz scalar $S_{LL}$, a transverse Lorentz vector $S_{LT}^\mu$ with two independent components $S_{LT}^x$ and $S_{LT}^y$, and a transverse Lorentz tensor $S_{TT}^{\mu\nu}$ with two independent components $S_{TT}^{xx} = -S_{TT}^{yy}$ and $S_{TT}^{xy}=S_{TT}^{yx}$~\cite{Bacchetta:2000jk,Chen:2016moq}.
Since $S_{LL}$ is a Lorentz scalar, there are nine $S_{LL}$-dependent BLTs in analog with the unpolarized part.
They are given by
\begin{align}
& h_{LLi}^{S\mu\nu}=S_{LL} h^{S\mu\nu}_{Ui}, \\
& \tilde h_{LLi}^{S\mu\nu}=S_{LL} \tilde h^{S\mu\nu}_{Ui}, \label{eq:hsLL} \\
& h_{LL}^{A\mu\nu}=S_{LL} h^{A\mu\nu}_{U}, \\
& \tilde h_{LLi}^{A\mu\nu}=S_{LL} \tilde h^{A\mu\nu}_{Ui}. \label{eq:thsLL}
\end{align}
The $S_{LT}^\mu$ dependent BLTs are given by
\begin{align}
& h^{S\mu\nu}_{LTi} =\Big\{ (p_{h\perp}\cdot S_{LT}) h^{S\mu\nu}_{Ui}, ~\varepsilon_\perp^{p_{h\perp} S_{LT}} \tilde h^{S\mu\nu}_{Uj}\Big\}, \label{eq:hsLT}\\
& \tilde h^{S\mu\nu}_{LTi} =\Big\{(p_{h\perp}\cdot S_{LT})\tilde h^{S\mu\nu}_{Ui}, ~ \varepsilon_\perp^{p_{h\perp} S_{LT}} h^{S\mu\nu}_{Uj} \Big\}, \label{eq:thsLT}\\
& h^{A\mu\nu}_{LTi} =\Big\{(p_{h\perp}\cdot S_{LT}) h^{A\mu\nu}_{U},~ \varepsilon_\perp^{p_{h\perp} S_{LT}}  \tilde h^{A\mu\nu}_{Uj}\Big\},\label{eq:haLT}\\
& \tilde h^{A\mu\nu}_{LTi} =\Big\{(p_{h\perp}\cdot S_{LT})\tilde h^{A\mu\nu}_{Ui},~ \varepsilon_\perp^{p_{h\perp} S_{LT}} h^{A\mu\nu}_{U}\Big\}. \label{eq:thaLT}
\end{align}
For the $S_{TT}^{\mu\nu}$ dependent part, we have
\begin{align}
& h^{S\mu\nu}_{TTi} =\Big\{ S_{TT}^{p_h p_h} h^{S\mu\nu}_{Ui}, ~\tilde S_{TT}^{p_h p_h}  \tilde h^{S\mu\nu}_{Uj}\Big\}, \label{eq:hsTT}\\
& \tilde h^{S\mu\nu}_{TTi} =\Big\{S_{TT}^{p_h p_h} \tilde h^{S\mu\nu}_{Ui}, ~ \tilde S_{TT}^{ p_h p_h} h^{S\mu\nu}_{Uj} \Big\}, \label{eq:thsTT}\\
& h^{A\mu\nu}_{TTi} =\Big\{S_{TT}^{p_h p_h}  h^{A\mu\nu}_{U},~ \tilde S_{TT}^{p_h p_h }  \tilde h^{A\mu\nu}_{Uj}\Big\}, \label{eq:haTT}\\
& \tilde h^{A,\mu\nu}_{TTi} =\Big\{S_{TT}^{p_h p_h} \tilde h^{A\mu\nu}_{Ui},~ \tilde S_{TT}^{ p_h p_h } h^{A\mu\nu}_{U}\Big\},\label{eq:thaTT}
\end{align}
where $S_{TT}^{p_h p_h} = p_{h\perp\alpha} p_{h\perp\beta} S_{TT}^{\alpha\beta}$ and $\tilde S_{TT}^{p_h p_h} = - \tilde p_{h\perp\alpha} p_{h\perp\beta} S_{TT}^{\alpha\beta}$ with $\tilde p_{h\perp}^\alpha = \varepsilon_\perp^{\alpha\mu} p_{h\perp\mu}$.

Substitute the hadronic tensors in Eqs. (\ref{eq:WSuv}) and (\ref{eq:WAuv}) into Eq. (\ref{eq:Xsec}), and after making Lorentz contractions with the leptonic tensor, we will obtain the general form for the cross section.
To be explicit, we first parametrize the components of the transverse polarization vector and tensor as
\begin{align}
& S_T^\mu = |S_T| \left( 0,0, \cos\phi_S, \sin\phi_S \right),\\
& S_{LT}^x = |S_{LT}| \cos\phi_{LT}, \\
& S_{LT}^y = |S_{LT}| \sin\phi_{LT}, \\
& S_{TT}^{xx} = |S_{TT}| \cos2\phi_{TT}, \\
& S_{TT}^{xy} = |S_{TT}| \sin2\phi_{TT}.
\end{align}
Then, the cross section can be divided into six parts according to the polarization states of the vector meson.
It is given by
\begin{widetext}
\begin{align}
\frac{d\sigma}{dx dy dz_h d\psi d^2 p_{h\perp}} &= \frac{\alpha_{\rm em}^2}{x y Q^2} \bigl( \mathcal{W}_{U} + \lambda \mathcal{W}_{L} + |S_T|\mathcal{W}_{T} + S_{LL}\mathcal{W}_{LL} + |S_{LT}|\mathcal{W}_{LT} + |S_{TT}|\mathcal{W}_{TT} \bigr),
\label{eq:Xsec-SFs}
\end{align}
where the subscripts of ${\cal W}$ denote the polarization states of the vector meson.
The explicit expressions for each part are calculated to be
\begin{align}
{\cal W}_{U} &= A(y) W_{U}^T + E(y) W_{U}^L + B(y)\left( \sin\phi \tilde W_{U1}^{\sin\phi} + \cos\phi W_{U1}^{\cos\phi} \right) + E(y)\left( \sin2\phi \tilde W_{U}^{\sin2\phi} + \cos2\phi W_{U}^{\cos2\phi} \right)  \nonumber\\
& + C(y) W_{U} + D(y) \left( \sin\phi \tilde W_{U2}^{\sin\phi} + \cos\phi W_{U2}^{\cos\phi} \right), \\
{\cal W}_{L} &= A(y) \tilde W_{L}^T + E(y) \tilde W_{L}^L + B(y)\left( \sin\phi W_{L1}^{\sin\phi} + \cos\phi \tilde W_{L1}^{\cos\phi} \right) + E(y)\left( \sin2\phi W_{L}^{\sin2\phi} + \cos2\phi \tilde W_{L}^{\cos2\phi} \right)  \nonumber\\
& + C(y) \tilde W_{L} + D(y) \left( \sin\phi W_{L2}^{\sin\phi} + \cos\phi \tilde W_{L2}^{\cos\phi} \right), \\
{\cal W}_{T} &= \sin\phi_S \left[ B(y) W_{T1}^{\sin\phi_S} + D(y) W_{T2}^{\sin\phi_S} \right]
+ \sin(\phi+\phi_S) E(y) W_{T}^{\sin(\phi+\phi_S)} \nonumber\\
&+ \sin(\phi-\phi_S) \left[ A(y) W_{T}^{T,\sin(\phi-\phi_S)} + E(y) W_{T}^{L,\sin(\phi-\phi_S)} + C(y) W_{T}^{\sin(\phi-\phi_S)} \right] \nonumber\\
&+ \sin(2\phi-\phi_S) \left[ B(y) W_{T1}^{\sin(2\phi-\phi_S)} + D(y) W_{T2}^{\sin(2\phi-\phi_S)} \right]
+ \sin(3\phi-\phi_S) E(y) W_{T}^{\sin(3\phi-\phi_S)} \nonumber\\
&+ \cos\phi_S \left[ B(y) \tilde W_{T1}^{\cos\phi_S} + D(y) \tilde W_{T2}^{\cos\phi_S} \right]
+ \cos(\phi+\phi_S) E(y) \tilde W_{T}^{\cos(\phi+\phi_S)} \nonumber\\
&+ \cos(\phi-\phi_S) \left[ A(y) \tilde W_{T}^{T,\cos(\phi-\phi_S)} + E(y) \tilde W_{T}^{L,\cos(\phi-\phi_S)} + C(y) \tilde W_{T}^{\cos(\phi-\phi_S)} \right] \nonumber\\
&+ \cos(2\phi-\phi_S) \left[ B(y) \tilde W_{T1}^{\cos(2\phi-\phi_S)} + D(y) \tilde W_{T2}^{\cos(2\phi-\phi_S)} \right]
+ \cos(3\phi-\phi_S) E(y) \tilde W_{T}^{\cos(3\phi-\phi_S)}, \\
{\cal W}_{LL} &= A(y) W_{LL}^T + E(y) W_{LL}^L + B(y)\left( \sin\phi \tilde W_{LL1}^{\sin\phi} + \cos\phi W_{LL1}^{\cos\phi} \right) + E(y)\left( \sin2\phi \tilde W_{LL}^{\sin2\phi} + \cos2\phi W_{LL}^{\cos2\phi} \right)  \nonumber\\
& + C(y) W_{LL} + D(y) \left( \sin\phi \tilde W_{LL2}^{\sin\phi} + \cos\phi W_{LL2}^{\cos\phi} \right), \\
{\cal W}_{LT} &= \sin\phi_{LT} \left[ B(y) \tilde W_{LT1}^{\sin\phi_{LT}} + D(y) \tilde W_{LT2}^{\sin\phi_{LT}} \right]
+ \sin(\phi+\phi_{LT}) E(y) \tilde W_{LT}^{\sin(\phi+\phi_{LT})} \nonumber\\
&+ \sin(\phi-\phi_{LT}) \left[ A(y) \tilde W_{LT}^{T,\sin(\phi-\phi_{LT})} + E(y) \tilde W_{LT}^{L,\sin(\phi-\phi_{LT})} + C(y) \tilde W_{LT}^{\sin(\phi-\phi_{LT})} \right] \nonumber\\
&+ \sin(2\phi-\phi_{LT}) \left[ B(y) \tilde W_{LT1}^{\sin(2\phi-\phi_{LT})} + D(y) \tilde W_{LT2}^{\sin(2\phi-\phi_{LT})} \right]
+ \sin(3\phi-\phi_{LT}) E(y) \tilde W_{LT}^{\sin(3\phi-\phi_{LT})} \nonumber\\
&+ \cos\phi_{LT} \left[ B(y) W_{LT1}^{\cos\phi_{LT}} + D(y) W_{LT2}^{\cos\phi_{LT}} \right]
+ \cos(\phi+\phi_{LT}) E(y) W_{LT}^{\cos(\phi+\phi_{LT})} \nonumber\\
&+ \cos(\phi-\phi_{LT}) \left[ A(y) W_{LT}^{T,\cos(\phi-\phi_{LT})} + E(y) W_{LT}^{L,\cos(\phi-\phi_{LT})} + C(y) W_{LT}^{\cos(\phi-\phi_{LT})} \right] \nonumber\\
&+ \cos(2\phi-\phi_{LT}) \left[ B(y) W_{LT1}^{\cos(2\phi-\phi_{LT})} + D(y) W_{LT2}^{\cos(2\phi-\phi_{LT})} \right]
+ \cos(3\phi-\phi_{LT}) E(y) W_{LT}^{\cos(3\phi-\phi_{LT})}, \\
{\cal W}_{TT} &= \sin(\phi-2\phi_{TT}) \left[ B(y) \tilde W_{TT1}^{\sin(\phi-2\phi_{TT})} + D(y) \tilde W_{TT2}^{\sin(\phi-2\phi_{TT})} \right]
+ \sin2\phi_{TT} E(y) \tilde W_{TT}^{\sin2\phi_{TT}} \nonumber\\
&+ \sin(2\phi-2\phi_{TT}) \left[ A(y) \tilde W_{TT}^{T,\sin(2\phi-2\phi_{TT})} + E(y) \tilde W_{TT}^{L,\sin(2\phi-2\phi_{TT})} + C(y) \tilde W_{TT}^{\sin(2\phi-2\phi_{TT})} \right] \nonumber\\
&+ \sin(3\phi-2\phi_{TT}) \left[ B(y) \tilde W_{TT1}^{\sin(3\phi-2\phi_{TT})} + D(y) \tilde W_{TT2}^{\sin(3\phi-2\phi_{TT})} \right]
+ \sin(4\phi-2\phi_{TT}) E(y) \tilde W_{TT}^{\sin(4\phi-2\phi_{TT})} \nonumber\\
&+ \cos(\phi-2\phi_{TT}) \left[ B(y) W_{TT1}^{\cos(\phi-2\phi_{TT})} + D(y) W_{TT2}^{\cos(\phi-2\phi_{TT})} \right]
+ \cos2\phi_{TT} E(y) W_{TT}^{\cos2\phi_{TT}} \nonumber\\
&+ \cos(2\phi-2\phi_{TT}) \left[ A(y) W_{TT}^{T,\cos(2\phi-2\phi_{TT})} + E(y) W_{TT}^{L,\cos(2\phi-2\phi_{TT})} + C(y) W_{TT}^{\cos(2\phi-2\phi_{TT})} \right] \nonumber\\
&+ \cos(3\phi-2\phi_{TT}) \left[ B(y) W_{TT1}^{\cos(3\phi-2\phi_{TT})} + D(y) W_{TT2}^{\cos(3\phi-2\phi_{TT})} \right]
+ \cos(4\phi-2\phi_{TT}) E(y) W_{TT}^{\cos(4\phi-2\phi_{TT})}.
\label{eq:SF}
\end{align}
\end{widetext}
We have defined the following functions of $y$ for simplicity:
\begin{align}
& A(y) = y^2-2y+2, \nonumber\\
& B(y) = 2(2-y)\sqrt{1-y}, \nonumber\\
& C(y) = y(2-y), \nonumber\\
& D(y) = 2y\sqrt{1-y}, \nonumber\\
& E(y) = 2(1-y).
\end{align}

There are 81 structure functions in total, which is exactly the number of the corresponding 81 independent BLTs for the hadronic tensor.
Among all the structure functions, 39 of them, i.e., the $\tilde W$'s, correspond to the space reflection odd part of the cross section and the remaining 42 correspond to the space reflection even part.
We also note that there are 45 structure functions that depend on the tensor polarizations of the vector meson.
It can be checked that the cross section for $e^-N \to e^- VX$ given in Eqs. (\ref{eq:Xsec-SFs})-(\ref{eq:SF}) has the same form in terms of the azimuthal angle dependence as that for $e^+e^- \to Z^0 \to V\pi X$ given in~\cite{Chen:2016moq}.
This is because these two processes share the same set of BLTs when constructing the general form of the hadronic tensors.

\section{Parton model calculations}
\label{sec:HT}
\subsection{The hadronic tensor results in terms of TMD PDFs and FFs}
We now calculate the hadronic tensor using the QCD parton model.
At the leading order of perturbative quantum chromodynamics and the leading twist, the hadronic tensor is given by
\begin{align}
W_{ZZ}^{\mu\nu}(q,p_N,&p_h,S) =\int d^4 k_i d^4 k_f \delta^4(q+k_i-k_f) \nonumber\\
& \times {\rm Tr} \left[ \hat \Phi(k_i,p_N) \Gamma^\mu \hat \Xi(k_f,p_h,S_h) \Gamma^\nu \right], 
\end{align}
where $\hat \Phi$ and $\hat \Xi$ are the parton correlators related to the parton distribution and the fragmentation process.
They are defined as~\cite{Bacchetta:2006tn}
\begin{align}
& \hat \Phi_{ij}(k_i,p_N) = \int \frac{d^4\xi}{(2\pi)^4} e^{i\xi\cdot k_i} \langle p_N | \bar\psi_{j}(0) \psi_{i}(\xi) | p_N \rangle, \\
& \hat \Xi_{ij}(k_f,p_h,S) = \sum_X \int \frac{d^4\eta}{(2\pi)^4} e^{-i\eta\cdot k_f } \langle 0 | \psi_{i}(0) |p_h,S;X\rangle \nonumber\\
&\hspace{4.2cm}\times \langle p_h,S;X | \bar\psi_{j}(\eta) |0\rangle.
\end{align}
We have suppressed the gauge links for short notations.
Taking approximations for the momenta in the $\delta$ function of the hadronic tensor, i.e., $k_i^-\approx0$ and $k_f^+\approx0$, and after integration, we get
\begin{align}
W_{ZZ}^{\mu\nu}
&= 2z_h\int d^2 k_{iT} d^2 k_{fT} \delta^2(\vec k_{iT}-\vec k_{fT} - \vec p_{h\perp}/z_h) \nonumber\\
&\qquad \times{\rm Tr} \left[ \hat \Phi(x,k_{iT}) \Gamma^\mu \hat \Xi(z_h,k_{fT},S) \Gamma^\nu \right], 
\label{eq:WZZ}
\end{align}
where the TMD parton correlators are defined by
\begin{align}
& \hat \Phi_{ij}(x,k_{iT}) = \int \frac{d\xi^- d^2\xi_T}{(2\pi)^3} e^{ixp_N^+\xi^- + i k_{iT}\cdot \xi_T} \nonumber\\
&\hspace{3cm} \times \langle p_N | \bar\psi_{i}(0) \psi_{j}(\xi^-,\xi_T) | p_N \rangle, \label{eq:Phi} \\
& \hat \Xi_{ij}(z_h,k_{fT},S) = \sum_X \int \frac{d\eta^+ d^2\eta_T}{2z_h(2\pi)^3} e^{-i\eta^+p_h^-/z_h - i \eta_T \cdot k_{fT} } \nonumber\\
&\qquad \times \langle 0 | \psi_{i}(0) |p_h,S;X\rangle \langle p_h,S;X | \bar\psi_{j}(\eta^+,\eta_T) |0\rangle.
\label{eq:Xi}
\end{align}

The parton correlators are $4\times4$ matrices in Dirac space.
They can be expanded under the basis of Gamma matrices, i.e.,
\begin{align}
& \hat \Phi = \frac{1}{2}\left( \Phi_\alpha \gamma^\alpha - \tilde \Phi_\alpha \gamma_5\gamma^\alpha + \Phi_{\alpha\beta} i\sigma^{\alpha\beta}\gamma_5 \right) + \cdots,
\label{eq:decPhi} \\
& \hat \Xi = \frac{1}{2}\left( \Xi_\alpha \gamma^\alpha + \tilde \Xi_\alpha \gamma_5\gamma^\alpha + \Xi_{\alpha\beta} i\sigma^{\alpha\beta}\gamma_5 \right) + \cdots,
\label{eq:decXi}
\end{align}
where $\cdots$ denotes terms irrelevant for the leading twist.
The $\Phi_\alpha$'s, etc., are expanding coefficients or are called correlation functions.
We have omitted the arguments of the correlators and correlation functions for short notations.
For an unpolarized nucleon, the Lorentz decomposition of the correlation functions only generate two TMD PDFs at the leading twist~\cite{Goeke:2005hb}.
We have
\begin{align}
& \Phi^\alpha = \bar n^\alpha f_1(x,k_{iT}) + \cdots, \label{eq:f1}\\
& \Phi^{\alpha\beta} = - \frac{1}{M_N} \bar n^\alpha \tilde k_{iT}^\beta h_1^\perp(x,k_{iT}) + \cdots. \label{eq:h1perp}
\end{align}
The first term defines the number density distribution, and $h_1^\perp$ is the Boer-Mulders function which is chiral-odd.

Similarly, for the fragmentation part, the correlation functions are decomposed as~\cite{Chen:2016moq}
\begin{align}
& \Xi^\alpha = n^\alpha \Big( D_1 + \frac{\tilde k_{fT} \cdot S_{T}}{M_h} D_{1T}^\perp + S_{LL} D_{1LL} \nonumber\\
&\qquad\quad + \frac{k_{fT}\cdot S_{LT}}{M_h} D_{1LT}^\perp + \frac{S_{TT}^{k_f k_f}}{M_h^2} D_{1TT}^\perp \Big)+ \cdots, \label{eq:Xia}\\
& \tilde \Xi^\alpha = n^\alpha \Big( \lambda_h G_{1L} + \frac{k_{fT} \cdot S_{T}}{M_h} G_{1T}^\perp \nonumber\\
&\qquad\quad + \frac{\tilde k_{fT} \cdot S_{LT}}{M_h} G_{1LT}^\perp - \frac{\tilde S_{TT}^{k_{f} k_{f}}}{M_h^2} G_{1TT}^\perp \Big)+ \cdots, \label{eq:tXia}\\
& \Xi^{\alpha\beta} = -\frac{n^{[\alpha} \tilde k_{fT}^{\beta]}}{M_h} \Big( H_1^\perp + S_{LL} H_{1LL}^\perp \Big) + n^{[\alpha} S_{T}^{\beta]} H_{1T} \nonumber\\
&\qquad\quad + \frac{n^{[\alpha} k_{fT}^{\beta]}}{M_h} \Big( \lambda_h H_{1L}^\perp + \frac{k_{fT}\cdot S_{T}}{M_h} H_{1T}^\perp \Big) \nonumber\\
&\qquad\quad - n^{[\alpha} \tilde S_{LT}^{\beta]} H_{1LT} - \frac{n^{[\alpha} \tilde S_{TT}^{k_f \beta]}}{M_h} H_{1TT}^{\prime\perp} \nonumber\\
&\qquad\quad - \frac{n^{[\alpha} \tilde k_{fT}^{\beta]}}{M_h} \Big(\frac{k_{fT} \cdot S_{LT}}{M_h} H_{1LT}^\perp + \frac{S_{TT}^{k_f k_f}}{M_h^2} H_{1TT}^\perp \Big) + \cdots. \label{eq:Xiab}
\end{align}
We have omitted the arguments of $z_h$ and $k_{fT}$ for the TMD FFs for short notations.
There are 18 TMD FFs at the leading twist, ten of them are tensor polarization dependent.

Substituting the decomposition of the parton correlators of Eqs.~(\ref{eq:decPhi})-(\ref{eq:Xiab}) into the hadronic tensor in Eq.~(\ref{eq:WZZ}), by carrying out the traces we can obtain the results for the hadronic tensor.
The relevant traces we need are
\begin{widetext}
\begin{align}
& {\rm Tr} [ \slashed{\bar n} \Gamma^\mu \slashed{n} \Gamma^\nu ] =
{\rm Tr} [ \gamma_5 \slashed{\bar n} \Gamma^\mu \gamma_5 \slashed{n} \Gamma^\nu ] = -4c_1^q g_\perp^{\mu\nu} - 4i c_3^q \varepsilon_\perp^{\mu\nu}, \\
& {\rm Tr} [ \slashed{\bar n} \Gamma^\mu \gamma_5 \slashed{n} \Gamma^\nu ] = {\rm Tr} [ \gamma_5\slashed{\bar n} \Gamma^\mu \slashed{n} \Gamma^\nu ] =
4c_3^q g_\perp^{\mu\nu} + 4i c_1^q \varepsilon_\perp^{\mu\nu}, \\
& {\rm Tr} \left[ i\sigma^{-\alpha} \gamma_5 \Gamma^\mu i\sigma^{+\beta} \gamma_5 \Gamma^\nu \right] 
= 4c_2^q \Big( g_\perp^{\mu\nu} g_\perp^{\alpha\beta} - g^{\mu\alpha} g^{\nu\beta} - g^{\nu\alpha} g^{\mu\beta} \Big),
\end{align}
where $c_2^q = (c_V^q)^2 - (c_A^q)^2$ and the indices $\alpha$ and $\beta$ are both transverse.
To further simplify the expressions, we define a symmetric tensor
$\alpha_{T}^{\mu\nu}(a_T,b_T) = a_T^{\{\mu} b_T^{\nu\}} - (a_T\cdot b_T)g_\perp^{\mu\nu}$, and we also use the following notations for the combinations of FFs:
\begin{align}
{\cal D}_1 &\equiv D_1 + \frac{\tilde k_{fT} \cdot S_{T}}{M_h} D_{1T}^\perp + S_{LL} D_{1LL} + \frac{k_{fT}\cdot S_{LT}}{M_h} D_{1LT}^\perp + \frac{S_{TT}^{k_f k_f}}{M_h^2} D_{1TT}^\perp, \\
{\cal G}_1 &\equiv \lambda_h G_{1L} + \frac{k_{fT} \cdot S_{T}}{M_h} G_{1T}^\perp + \frac{\tilde k_{fT} \cdot S_{LT}}{M_h} G_{1LT}^\perp - \frac{\tilde S_{TT}^{k_{f} k_{f}}}{M_h^2} G_{1TT}^\perp, \\
{\cal H} &\equiv H_1^\perp  + S_{LL} H_{1LL}^\perp + \frac{k_{fT} \cdot S_{LT}}{M_h} H_{1LT}^\perp + \frac{S_{TT}^{k_f k_f}}{M_h^2} H_{1TT}^\perp.
\end{align}
Using these notations, we obtain the hadronic tensor given by
\begin{align}
W_{ZZ}^{\mu\nu} &= \frac{2z_h}{x} {\cal C} \Biggl\{ (c_3^q g_\perp^{\mu\nu} + ic_1^q \varepsilon_\perp^{\mu\nu}) f_1 {\cal G}_1  - (c_1^q g_\perp^{\mu\nu} + ic_3^q \varepsilon_\perp^{\mu\nu}) f_1 {\cal D}_1 \nonumber\\
&\qquad + 2 c_2^q h_1^\perp \biggl[ \frac{\alpha_T^{\mu\nu}(\tilde k_{iT}, S_{T})}{M_N} H_{1T} 
+ \frac{\alpha_T^{\mu\nu}(\tilde k_{iT}, k_{fT})}{M_N M_h} \biggl( \lambda_h H_{1L}^\perp + \frac{k_{fT}\cdot S_{T}}{M_h} H_{1T}^\perp\biggr) \nonumber\\
& \qquad\qquad + \frac{\alpha_T^{\mu\nu}(k_{iT}, S_{LT})}{M_N} H_{1LT} 
+ \frac{\alpha_T^{\mu\nu}(k_{iT}, S_{TT}^{k_f})}{M_N M_h} H_{1TT}^{\prime\perp} 
+ \frac{\alpha_T^{\mu\nu}(k_{iT}, k_{fT})}{M_N M_h} {\cal H} \biggr]
\Biggr\}.
\label{eq:W-result}
\end{align}
The convolution ${\cal C}[\cdots]$ is defined as
\begin{align}
{\cal C}[wfD] = x\int d^2 k_{iT} d^2 k_{fT} \delta^2(\vec k_{iT}-\vec k_{fT} - \vec p_{h\perp}/z_h) w f(x,k_{iT}) D(z_h, k_{fT}).
\end{align}
We see that the hadronic tensor is given by the convolution of the TMD PDFs and FFs.
There are 18 different convolution modules, in which half of them are related to the chiral-odd TMD PDFs and FFs.

\end{widetext}

\subsection{The cross section in the parton model}
Substituting the hadronic tensor Eq.~(\ref{eq:W-result}) into Eq.~(\ref{eq:Xsec}), and making Lorentz contractions with the leptonic tensor, we will obtain the cross section expressed by the TMD PDFs and FFs.
However, we notice that the indices in the hadronic tensor are taken by the convolution variables $k_{iT}$ and $k_{fT}$.
In order to compare with the structure function results, we need to do a transformation to rewrite these terms using physical observables, so that the azimuthal angle modulations will be explicit.
To this end, we define a unit perpendicular vector $\hat h^\mu = p_{h\perp}^\mu/|\vec p_{h\perp}|$ representing the direction of the transverse momentum of the vector meson.
Under the convolution, we have, from Lorentz covariance, e.g.,
\begin{align}
{\cal C}[ k_{iT}^{\mu} ~{\rm F}(k_{iT},k_{fT}, \hat h)] = A \hat h^\mu,
\end{align}
where ${\rm F}(k_{iT},k_{fT}, \hat h)$ is an arbitrary scalar function of $k_{iT}$, $k_{fT}$, and $\hat h$.
The coefficient $A$ can be obtained by contraction on both sides with the unit vector $\hat h_\mu$, i.e.,
\begin{align}
A = -{\cal C}[ (\hat h \cdot k_{iT}) ~{\rm F}(k_{iT},k_{fT}, \hat h)].
\end{align}
This means that, under the convolution, one can equivalently replace $k_{iT}^\mu$ with $-(\hat h \cdot k_{iT}) \hat h^\mu$.
For terms with two or more indices, we give the detailed derivations and expressions in Appendix~\ref{ap:1}.
These algebras can also be found with a compact form in Ref.~\cite{Tangerman:1994eh}.

When contracting the hadronic tensor with the leptonic tensor, we need the following basic contraction results, i.e.,
\begin{align}
& L_{ZZ}^{\mu\nu} \cdot (c_1^q g_{\perp\mu\nu} + ic_3^q \varepsilon_{\perp\mu\nu}) = -\frac{2Q^2}{y^2}  T_0^q(y), \\
& L_{ZZ}^{\mu\nu} \cdot (c_3^q g_{\perp\mu\nu} + ic_1^q \varepsilon_{\perp\mu\nu}) = -\frac{2Q^2}{y^2} \tilde T_1^q(y), \\
& L_{ZZ}^{\mu\nu}  \cdot \alpha_{T\mu\nu}(a_T, b_T) = c_1^e \frac{2Q^2 E(y)}{y^2}|\vec a_T||\vec b_T| \cos(\phi_a + \phi_b), \\
& L_{ZZ}^{\mu\nu}  \cdot \alpha_{T\mu\nu}(\tilde a_T, b_T) = -c_1^e \frac{2Q^2 E(y)}{y^2}|\vec a_T||\vec b_T| \sin(\phi_a + \phi_b),
\label{eq:contractions}
\end{align}
where we have defined
\begin{align}
 & T_0^q(y) = c_1^e c_1^q A(y) + c_3^e c_3^q C(y), \\
 & \tilde T_1^q(y) = c_1^e c_3^q A(y) + c_3^e c_1^q C(y).
\label{eq:T0T1}
\end{align}
We will also define and use the following dimensionless coefficients to simplify the results:
\begin{align}
\bar w_0 &= -\frac{k_{fT}^2}{M_h^2}, \\
w_0^\prime &= -\frac{k_{iT}\cdot k_{fT}}{M_N M_h}, \\
w_1 &= -\frac{\hat h \cdot k_{iT}}{M_N}, \\
\bar w_1 &= -\frac{\hat h \cdot k_{fT}}{M_h}, \\
w_2 &= \frac{2(\hat h\cdot k_{iT})(\hat h\cdot k_{fT}) + k_{iT}\cdot k_{fT}}{M_N M_h}, \\
w_2^\prime &= \frac{(\hat h\cdot k_{iT})(\hat h\cdot k_{fT}) + k_{iT}\cdot k_{fT}}{M_N M_h}, \\
\bar w_2 &= \frac{2(\hat h \cdot k_{fT})^2 + k_{fT}^2}{M_h^2}, \\
w_3 &= \frac{-1}{M_N M_h^2} \big[4(\hat h \cdot k_{fT})^2 (\hat h \cdot k_{iT}) + k_{fT}^2(\hat h \cdot k_{iT}) \nonumber\\
&\qquad\qquad + 2(\hat h \cdot k_{fT}) (k_{iT}\cdot k_{fT}) \big], \\
w_4 &= \frac{1}{M_N M_h^3} \big\{ k_{fT}^2 [ k_{iT}\cdot k_{fT} + 4 (\hat h \cdot k_{fT}) (\hat h \cdot k_{iT}) ]  \nonumber\\
&~~ + 4(\hat h \cdot k_{fT})^2 [ k_{iT} \cdot k_{fT} + 2 (\hat h \cdot k_{fT}) (\hat h \cdot k_{iT})] \big\}.
\end{align}
We divide the cross section into two parts, according to the chirality of the TMD PDFs or FFs involved, i.e.,
\begin{align}
d\sigma_{ZZ} = d\sigma_{ZZ}\vert_{\rm \chi-even} + d\sigma_{ZZ}\vert_{\rm \chi-odd}.
\end{align}
The chiral-even part is calculated as
\begin{align}
&\frac{d\sigma_{ZZ}\vert_{\rm \chi-even}}{dx dy dz_h d\psi d^2 p_{h\perp}} = \frac{\alpha_{\rm em}^2\chi}{xyQ^2} \nonumber\\
&\times \Bigl\{ T_0^q(y) {\cal C} \Bigl[ f_1 (D_1 + S_{LL} D_{1LL}) \nonumber\\
&\qquad\qquad + |S_T| \sin(\phi-\phi_S) \bar w_1 f_1 D_{1T}^\perp \nonumber\\
&\qquad\qquad - |S_{LT}| \cos(\phi-\phi_{LT}) \bar w_1 f_1 D_{1LT}^\perp \nonumber\\
&\qquad\qquad + |S_{TT}| \cos(2\phi-2\phi_{TT}) \bar w_2 f_1 D_{1TT}^\perp \Bigr] \nonumber\\
&\quad - \tilde T_1^q(y) {\cal C}\Bigl[ \lambda f_1 G_{1L} - |S_T| \cos(\phi-\phi_S) \bar w_1 f_1 G_{1T}^\perp \nonumber\\
&\qquad\qquad + |S_{LT}| \sin(\phi-\phi_{LT}) \bar w_1 f_1 G_{1LT}^\perp \nonumber\\
&\qquad\qquad - |S_{TT}| \sin(2\phi-2\phi_{TT}) \bar w_2 f_1 G_{1TT}^\perp \Bigr]\Bigr\}.
\label{eq:Xsec-even}
\end{align}
The chiral-odd part is calculated as
\begin{align}
&\frac{d\sigma_{ZZ}\vert_{\rm \chi-odd}}{dx dy dz_h d\psi d^2 p_{h\perp}} = \frac{\alpha_{\rm em}^2\chi}{xyQ^2} c_1^e c_2^q E(y) \nonumber\\
&\times {\cal C}\Bigl\{
\cos2\phi~ 2w_2 h_1^\perp \left( H_1^\perp + S_{LL} H_{1LL}^\perp \right) \nonumber\\
& - \lambda \sin2\phi~ 2w_2 h_1^\perp H_{1L}^\perp - |S_T| \sin(\phi + \phi_S)~ 2w_1 h_1^\perp H_{1T} \nonumber\\
& + |S_T| \bigl[ \bar w_0 w_1 \sin(\phi + \phi_{S}) + w_3 \sin(3\phi - \phi_{S}) \bigr] h_1^\perp H_{1T}^\perp \nonumber\\
& + |S_{LT}|  \cos(\phi + \phi_{LT})~ 2w_1 h_1^\perp H_{1LT} \nonumber\\
& - |S_{LT}| [ w_3 \cos(3\phi - \phi_{LT}) + \bar w_0 w_1 \cos(\phi + \phi_{LT}) ] h_1^\perp H_{1LT}^\perp \nonumber\\
& + |S_{TT}| \bigl[ w_4 \cos(4\phi - 2\phi_{TT}) + \bar w_0 w_0^\prime \cos2\phi_{TT} \bigr] h_1^\perp H_{1TT}^\perp \nonumber\\
& - |S_{TT}| \cos2\phi_{TT}~ 2w_0^\prime h_1^\perp H_{1TT}^{\prime\perp} \Bigr\}.
\label{eq:Xsec-odd}
\end{align}

It is straightforward to obtain the EM and the interference contributions by doing replacements for the electroweak coefficients, e.g.,
$c_2^q \to 1$ and $c_V^q$ for the EM and the interference parts, respectively.
To further unify the notations, we define $T_{0,r}^q(y)$'s and $\tilde T_{1,r}^q(y)$'s with $r=ZZ$, $\gamma Z$, and $\gamma\gamma$.
For the weak interaction part, we have $T_{0,ZZ}^q(y) = T_{0}^q(y)$ and $\tilde T_{1,ZZ}^q(y) = \tilde T_{1}^q(y)$.
For $\gamma Z$ and $\gamma\gamma$ parts, we have
\begin{align}
& T_{0,\gamma Z}^q(y) = c_V^e c_V^q A(y) + c_A^e c_A^q C(y), \\
& \tilde T_{1,\gamma Z}^q(y) = c_V^e c_A^q A(y) + c_A^e c_V^q C(y), \\
& T_{0,\gamma\gamma}^q(y) = A(y), \\
& \tilde T_{1,\gamma\gamma}^q(y) = 0.
\end{align}
For simplicity, we will not show the total cross section explicitly.
Instead, we give the explicit parton model results for the structure functions in the next section.

\section{The structure functions results and the spin alignment}
\label{sec:Results}
\subsection{Parton model results of the structure functions}
By comparing the cross section given by the structure functions in Eq. (\ref{eq:Xsec-SFs}) and the parton model results in Eqs. (\ref{eq:Xsec-even}) and (\ref{eq:Xsec-odd}),
we get the structure functions results in terms of the convolution of the TMD PDFs and FFs.
We include all the contributions from $\gamma\gamma$, $ZZ$, and $\gamma Z$ channels, i.e., Eq. (\ref{eq:TotalXsec}).
To simplify the expressions, we define the following electroweak coefficients:
\begin{align}
& c_{11}^{\rm ew} = c_1^e c_1^q \chi + c_V^e c_V^q \chi_{\rm int} + e_q^2, \\
& c_{12}^{\rm ew} = c_1^e c_2^q \chi + c_V^e c_V^q \chi_{\rm int} + e_q^2, \\
& c_{13}^{\rm ew} = c_1^e c_3^q \chi + c_V^e c_A^q \chi_{\rm int}, \\
& c_{31}^{\rm ew} = c_3^e c_1^q \chi + c_A^e c_V^q \chi_{\rm int}, \\
& c_{33}^{\rm ew} = c_3^e c_3^q \chi + c_A^e c_A^q \chi_{\rm int}.
\end{align}
We give the structure functions results in the following:
\begin{align}
& W_U^T = c_{11}^{\rm ew} {\cal C} [f_1 D_1], \\
& W_U = c_{33}^{\rm ew} {\cal C} [f_1 D_1], \\
& W_{U}^{\cos2\phi} = 2 c_{12}^{\rm ew} {\cal C} [w_2 h_1^\perp H_1^\perp],\\
& \tilde W_L^T = -c_{13}^{\rm ew} {\cal C} [f_1 G_{1L}], \\
& \tilde W_L = -c_{31}^{\rm ew} {\cal C} [f_1 G_{1L}], \\
& W_{L}^{\sin2\phi} = -2c_{12}^{\rm ew} {\cal C} [w_2 h_1^\perp H_{1L}^\perp], \\
& W_{T}^{T,\sin(\phi-\phi_S)} = c_{11}^{\rm ew} {\cal C} [\bar w_1 f_1 D_{1T}^\perp], \\
& W_{T}^{\sin(\phi-\phi_S)} = c_{33}^{\rm ew} {\cal C} [\bar w_1 f_1 D_{1T}^\perp], \\
& \tilde W_{T}^{T,\cos(\phi-\phi_S)} = c_{13}^{\rm ew} {\cal C} [\bar w_1 f_1 G_{1T}^\perp], \\
& \tilde W_{T}^{\cos(\phi-\phi_S)} = c_{31}^{\rm ew} {\cal C} [\bar w_1 f_1 G_{1T}^\perp], \\
& W_{T}^{\sin(\phi+\phi_S)} = c_{12}^{\rm ew} {\cal C} [w_1 h_1^\perp (\bar w_0 H_{1T}^\perp - 2H_{1T})], \\
& W_{T}^{\sin(3\phi-\phi_S)} = c_{12}^{\rm ew} {\cal C} [w_3 h_1^\perp H_{1T}^\perp], \\
& W_{LL}^T = c_{11}^{\rm ew} {\cal C} [f_1 D_{1LL}], \\
& W_{LL} = c_{33}^{\rm ew} {\cal C} [f_1 D_{1LL}], \\
& W_{LL}^{\cos2\phi} = 2 c_{12}^{\rm ew} {\cal C} [w_2 h_1^\perp H_{1LL}^\perp],\\
& W_{LT}^{T,\cos(\phi-\phi_{LT})} = -c_{11}^{\rm ew} {\cal C} [\bar w_1 f_1 D_{1LT}^\perp], \\
& W_{LT}^{\cos(\phi-\phi_{LT})} = -c_{33}^{\rm ew} {\cal C} [\bar w_1 f_1 D_{1LT}^\perp], \\
& \tilde W_{LT}^{T,\sin(\phi-\phi_{LT})} = -c_{13}^{\rm ew} {\cal C} [\bar w_1 f_1 G_{1LT}^\perp], \\
& \tilde W_{LT}^{\sin(\phi-\phi_{LT})} = -c_{31}^{\rm ew} {\cal C} [\bar w_1 f_1 G_{1LT}^\perp], \\
& W_{LT}^{\cos(\phi+\phi_{LT})} = c_{12}^{\rm ew} {\cal C} [w_1 h_1^\perp (2H_{1LT} - \bar w_0 H_{1LT}^\perp)], \\
& W_{LT}^{\cos(3\phi-\phi_{LT})} = -c_{12}^{\rm ew} {\cal C} [w_3 h_1^\perp H_{1LT}^\perp], \\
& W_{TT}^{T,\cos(2\phi-2\phi_{TT})} = c_{11}^{\rm ew} {\cal C} [\bar w_2 f_1 D_{1TT}^\perp], \\
& W_{TT}^{\cos(2\phi-2\phi_{TT})} = c_{33}^{\rm ew} {\cal C} [\bar w_2 f_1 D_{1TT}^\perp], \\
& \tilde W_{TT}^{T,\sin(2\phi-2\phi_{TT})} = c_{13}^{\rm ew} {\cal C} [\bar w_2 f_1 G_{1TT}^\perp], \\
& \tilde W_{TT}^{\sin(2\phi-2\phi_{TT})} = c_{31}^{\rm ew} {\cal C} [\bar w_2 f_1 G_{1TT}^\perp], \\
& W_{TT}^{\cos2\phi_{TT}} = c_{12}^{\rm ew} {\cal C} [w_0^\prime h_1^\perp (\bar w_0 H_{1TT}^\perp - 2H_{1TT}^{\prime\perp})], \\
& W_{TT}^{\cos(4\phi-2\phi_{TT})} = c_{12}^{\rm ew} {\cal C} [w_4 h_1^\perp H_{1TT}^\perp].
\end{align}

We see that there are 27 nonzero structure functions at the leading twist. 
Among these structure functions, 15 are related to the tensor polarizations of the vector meson.
The structure functions denoted by $\tilde W$'s are related to $c_{13}^{\rm ew}$ and $c_{31}^{\rm ew}$, which are labels of parity odd structure.
If we only consider the EM interaction, 14 terms that are associated with $c_{11}^{\rm ew}$ and $c_{12}^{\rm ew}$ will survive.
After reducing to EM interaction, it can be checked that the unpolarized and vector polarization dependent parts are consistent with the results given in, e.g.,~\cite{Bacchetta:2006tn, Yang:2016qsf}.
The other 13 structure functions (those related to $c_{13}^{\rm ew}$, $c_{31}^{\rm ew}$, and $c_{33}^{\rm ew}$) will vanish.
These 13 structure functions are generated by the weak interaction and the interference between the EM and weak interactions.

\subsection{The spin alignment of the vector meson}
Compared with the hyperon production, the tensor polarizations are unique for polarized vector meson production.
The tensor polarizations of the vector meson can be measured through the angular distribution of their decay products~\cite{OPAL:1997nwj}.
Among different components of the tensor polarizations, the spin alignment is perhaps the most interesting one that has been studied a lot.
The spin alignment $\rho_{00}^V$ is defined by the $00$ component of the spin density matrix in the helicity basis.
In terms of the differential cross section, it is given by~\cite{Wei:2013csa}
\begin{align}
\rho_{00}^{V} = \frac{d\sigma(\lambda = 0)}{\sum_{\lambda = \pm 1,0} d\sigma(\lambda)}.
\end{align}
For the helicity $\lambda = \pm 1$ states, $S_{LL} = 1/2$, while $S_{LL} = -1$ for the $\lambda = 0$ state, and all the other polarization components are zero.
Therefore, from the general form of the cross section in Eq.~(\ref{eq:Xsec-SFs}), we get
\begin{align}
\rho_{00}^V = \frac{1}{3} - \frac{{\cal W}_{LL}}{3{\cal W}_U}.
\end{align}
Substituting the parton model results of the structure functions, we obtain
\begin{align}
\rho_{00}^V &= \frac{1}{3}\Bigl\{ 1 - \Bigl[ \bigl( c_{11}^{\rm ew} A(y) + c_{33}^{\rm ew} C(y) \bigr) {\cal C}[f_1 D_{1LL}] + 2c_{12}^{\rm ew} E(y) \nonumber\\
& \times{\cal C}[w_2 h_1^\perp H_{1LL}^\perp] \cos2\phi \Bigr] {\Big /} 
\Bigl[ \bigl( c_{11}^{\rm ew} A(y) + c_{33}^{\rm ew} C(y) \bigr) {\cal C}[f_1 D_{1}] \nonumber\\
& + 2c_{12}^{\rm ew} E(y) {\cal C}[w_2 h_1^\perp H_{1}^\perp] \cos2\phi \Bigr] \Bigr\}.
\end{align}
We see that the spin alignment depends on both the chiral-even FF $D_{1LL}$ and the chiral-odd FF $H_{1LL}^\perp$.
It also depends on the azimuthal angle $\phi$ in general.
However, experimentally, it is much easier to measure the $\phi$ integrated spin alignment $\langle \rho_{00}^V \rangle$.
In this case, we have
\begin{align}
\langle \rho_{00}^V \rangle = \frac{1}{3} - \frac{\bigl( c_{11}^{\rm ew} A(y) + c_{33}^{\rm ew} C(y) \bigr) {\cal C}[f_1 D_{1LL}]}{3\bigl( c_{11}^{\rm ew} A(y) + c_{33}^{\rm ew} C(y) \bigr) {\cal C}[f_1 D_{1}]}.
\label{eq:SpinAl-int}
\end{align}
If we only consider the EM interaction, we can easily get the expression reduced from Eq.~(\ref{eq:SpinAl-int}) given by
\begin{align}
\langle \rho_{00}^V \rangle {\big \vert_{\rm em}} = \frac{1}{3} - \frac{ e_q^2 {\cal C}[f_1 D_{1LL}]}{3 e_q^2 {\cal C}[f_1 D_{1}]}
\label{eq:SpinAl-int-em}
\end{align}
that a summation over different quark flavors is implicit both in Eqs. (\ref{eq:SpinAl-int}) and (\ref{eq:SpinAl-int-em}).
We see that $\langle \rho_{00}^V \rangle {\big \vert_{\rm em}}$ is much simpler than $\langle \rho_{00}^V \rangle$ and independent of $y$. 
It is clear that the spin alignment of the vector meson is independent of the quark polarization and will deviate from $1/3$ if the $D_{1LL}$ FF is nonzero.

\begin{figure}[!htb]
\includegraphics[width=0.45\textwidth]{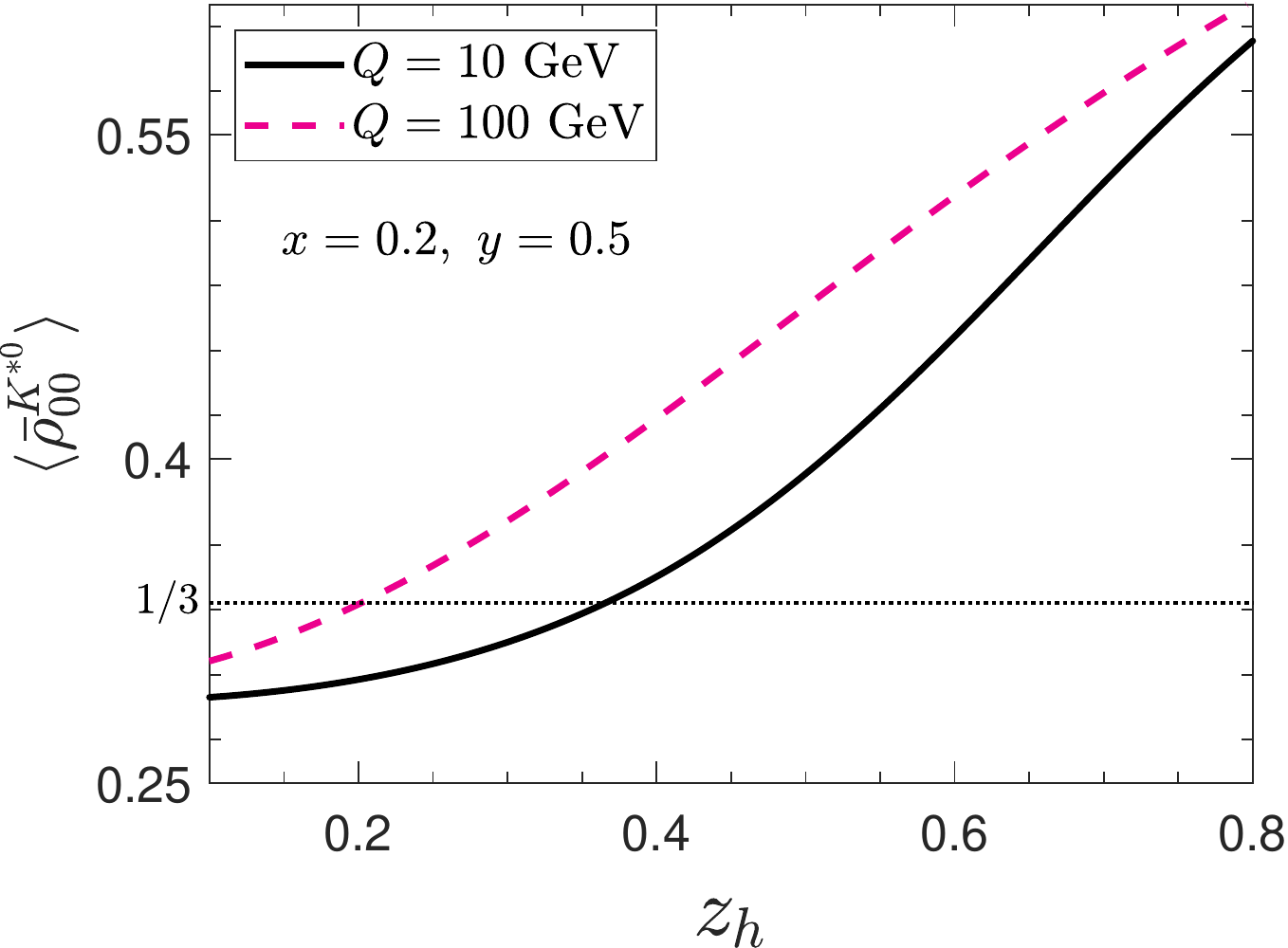}
\caption{A rough numerical estimate of the spin alignment for $K^{*0}$ production.}
\label{fig:SpinAl}
\end{figure}
We take the production of the $K^{*0}$ vector meson as an example to give a rough numerical estimate of the spin alignment.
For the corresponding TMD PDFs and FFs, we consider only light flavors and take the Gaussian ansatz, i.e.,
\begin{align}
& f_1(x, k_{iT}) = f_{1q}(x) \frac{1}{\pi \Delta_f^2} e^{-\vec k_{iT}^2/\Delta_f^2}, \label{eq:TMDf1} \\
& D_1(z_h, k_{fT}) = D_{1q}^{K^{*0}}(z_h) \frac{1}{\pi \Delta_D^2} e^{-\vec k_{fT}^2/\Delta_D^2}, \label{eq:TMDD1} \\
& D_{1LL}(z_h, k_{fT}) = D_{1LLq}^{K^{*0}}(z_h) \frac{1}{\pi \Delta_{LL}^2} e^{-\vec k_{fT}^2/\Delta_{LL}^2}. \label{eq:TMDD1LL}
\end{align}
They are factorized into a collinear distribution part and a Gaussian distribution part for the transverse momentum dependence.
It should be noted that the Gaussian widths, i.e., $\Delta_f$, $\Delta_D$, and $\Delta_{LL}$ can depend on different flavors in principle.
However, if we substitute Eqs.~(\ref{eq:TMDf1})-(\ref{eq:TMDD1LL}) into Eq.~(\ref{eq:SpinAl-int}), carry out the convolution integrals and further integrate over the transverse momentum, $p_{h\perp}$, of the produced vector meson, the Gaussian widths will cancel.
More explicitly, we get the $p_{h\perp}$ integrated (or averaged) spin alignment given by the collinear part of the TMD PDFs and FFs, i.e.,
\begin{align}
\langle \bar \rho_{00}^{K^{*0}} \rangle = \frac{1}{3} - \frac{\bigl[ c_{11}^{\rm ew} A(y) + c_{33}^{\rm ew} C(y) \bigr] f_{1q}(x) D_{1LLq}^{K^{*0}}(z_h)}{3\bigl[ c_{11}^{\rm ew} A(y) + c_{33}^{\rm ew} C(y) \bigr] f_{1q}(x) D_{1q}^{K^{*0}}(z_h)},
\label{eq:SpinAl-avr}
\end{align}
where a summation over quark flavors is also implicit in the numerator and the denominator of Eq. (\ref{eq:SpinAl-avr}).

We choose $x = 0.2$ and $y = 0.5$ as a typical value and show the rough numerical estimate of the spin alignment as a function of $z_h$ in Fig.~\ref{fig:SpinAl} for $Q = 10~{\rm GeV}$ and $100~{\rm GeV}$, respectively.
In the numerical calculation, we have taken the CT14 next-to-leading order PDFs~\cite{Dulat:2015mca} for $f_{1q}(x)$.
For the FFs $D_{1q}^{K^{*0}}(z_h)$ and $D_{1LLq}^{K^{*0}}(z_h)$, we use the parametrization results given in~\cite{Chen:2020pty}.
The factorization scales for the PDFs and FFs are set to $\mu_f = Q$.
It is clear to see that the spin alignment is deviated from $1/3$ at both low and high $Q$ values.
We also note that the spin alignment increases monotonically with $z_h$.
These properties may be checked in the SIDIS experiments such as JLab or EIC in the future.

\section{Summary}
\label{sec:Sum}
Semi-inclusive deep inelastic scattering is an important process for accessing the three-dimensional partonic structure of the nucleon and the hadronization mechanism.
We present a systematic calculation for $e^-N\to e^-VX$ with unpolarized electron and nucleon beams and polarized vector meson production at high energies.
We give a full kinematic analysis for this process by considering both the electromagnetic and the weak interactions that introduce the parity-violating effects. 
The results show that the cross sections are expressed by 81 structure functions.
Among all the structure functions, 39 correspond to the space reflection odd part of the cross section and 42 correspond to the space reflection even part.
We also carry out a parton model calculation for the process and show that there are 27 nonzero structure functions at the leading twist, in which 15 are related to the tensor polarizations of the vector meson and 13 are generated by the parity-violating effects.
The structure functions are given by the convolution of the corresponding TMD PDFs and FFs.
We also present the result for the spin alignment of the vector meson.
A rough numerical estimate is made, as an example, for the $K^{*0}$ spin alignment.
It gives us ways for accessing the tensor polarization dependent FFs by measuring the polarization of the vector meson.
Future experimental studies such as the EIC will provide us with better opportunities to study the nucleon structure and the hadronization mechanism as well as the polarization effects in detail.

\begin{acknowledgments}

This work was supported by the National Natural Science Foundation of China (Approvals No. 12005122 and No. 11947055) and Shandong Province Natural Science Foundation Grant No. ZR2020QA082.

\end{acknowledgments}

\appendix
\begin{widetext}
\section{Transformations for the convolution}
\label{ap:1}
For terms with two indices, according to Lorentz covariance, we have
\begin{align}
{\cal C}[ k_{iT}^{\mu} k_{fT}^\nu ~{\rm F}(k_{iT},k_{fT}, \hat h)] = B_1 \hat h^\mu \hat h^\nu + B_2 g_\perp^{\mu\nu}.
\end{align}
The coefficients $B_1$ and $B_2$ can be obtained by contracting with $\hat h_\mu \hat h_\nu$ and $g_{\perp\mu\nu}$ on both sides, i.e.,
\begin{align}
& {\cal C} [(\hat h \cdot k_{iT}) (\hat h \cdot k_{fT})~{\rm F}(k_{iT},k_{fT}, \hat h)] = B_1 - B_2, \\
& {\cal C} [(k_{iT} \cdot k_{fT})~{\rm F}(k_{iT},k_{fT}, \hat h)] = -B_1 + 2B_2.
\end{align}
One can solve
\begin{align}
& B_1 = {\cal C}\Big\{ \big[2(\hat h\cdot k_{iT})(\hat h\cdot k_{fT}) + k_{iT}\cdot k_{fT}\big] ~{\rm F}(k_{iT},k_{fT}, \hat h) \Big\},\\
& B_2 = {\cal C}\Big\{ \big[(\hat h\cdot k_{iT})(\hat h\cdot k_{fT}) + k_{iT}\cdot k_{fT}\big] ~{\rm F}(k_{iT},k_{fT}, \hat h) \Big\}.
\end{align}
For terms with three indices, we only need to consider the form of ${\cal C}[ k_{iT}^{\rho} k_{fT}^\mu k_{fT}^\nu ~{\rm F}(k_{iT},k_{fT}, \hat h)]$ in the calculation.
Note that the $\mu$ $\nu$ indices are symmetric in this case and can be expressed by three bases. We have
\begin{align}
{\cal C}[ k_{iT}^{\rho} k_{fT}^\mu k_{fT}^\nu ~{\rm F}(k_{iT},k_{fT}, \hat h)] = C_1 \hat h^\rho \hat h^\mu \hat h^\nu + C_2 \hat h^\rho g_\perp^{\mu\nu} + C_3 g_\perp^{\rho\{\mu} \hat h^{\nu\}}.
\end{align}
By contractions on both side with the three bases, one can also solve
\begin{align}
& C_1 = {\cal C}\Big\{ \big[-4(\hat h \cdot k_{fT})^2 (\hat h \cdot k_{iT}) - 2(\hat h \cdot k_{fT}) (k_{iT}\cdot k_{fT}) - k_{fT}^2(\hat h \cdot k_{iT})\big] ~{\rm F}(k_{iT},k_{fT}, \hat h) \Big\}, \\
& C_2 = {\cal C}\Big\{ \big[-(\hat h \cdot k_{fT})^2 (\hat h \cdot k_{iT})  - k_{fT}^2(\hat h \cdot k_{iT})\big] ~{\rm F}(k_{iT},k_{fT}, \hat h) \Big\}, \\
& C_3 = {\cal C}\Big\{ \big[-(\hat h \cdot k_{fT})^2 (\hat h \cdot k_{iT}) - (\hat h \cdot k_{fT}) (k_{iT}\cdot k_{fT})\big] ~{\rm F}(k_{iT},k_{fT}, \hat h) \Big\}.
\end{align}
For terms with four $k_T$'s, we need to consider the convolution with the form of ${\cal C}[k_{iT}^{\{\mu} k_{fT}^{\nu\}} k_{fT}^\rho k_{fT}^\sigma~{\rm F}(k_{iT},k_{fT}, \hat h)]$.
In this case, the $\mu$ $\nu$ and $\rho$ $\sigma$ indices are symmetric, respectively.
Thus, we have
\begin{align}
{\cal C}[k_{iT}^{\{\mu} k_{fT}^{\nu\}} k_{fT}^\rho k_{fT}^\sigma~{\rm F}(k_{iT},k_{fT}, \hat h)] &= D_1 \hat h^\mu \hat h^\nu \hat h^\rho \hat h^\sigma + D_2 \hat h^\mu \hat h^\nu g_\perp^{\rho\sigma} + D_3 \hat h^\rho \hat h^\sigma g_\perp^{\mu\nu} + D_4 \frac{1}{2}\bigl( \hat h^\rho \hat h^{\{\mu} g_\perp^{\nu\}\sigma} + \hat h^\sigma \hat h^{\{\mu} g_\perp^{\nu\}\rho} \bigr)\nonumber\\
& + D_5 g_\perp^{\mu\nu} g_\perp^{\rho\sigma}.
\end{align}
Also by contractions on both sides with the five bases, one can solve and get
\begin{align}
& D_1 = {\cal C}\Bigl\{ \bigl[ 2 k_{fT}^2 [ k_{iT}\cdot k_{fT} + 4 (\hat h \cdot k_{fT}) (k_{iT} \cdot k_{fT}) ]  +8 (\hat h \cdot k_{fT})^2 [ k_{iT} \cdot k_{fT} + 2 (\hat h \cdot k_{fT}) (\hat h \cdot k_{iT})] \bigr]~{\rm F}(k_{iT},k_{fT}, \hat h) \Bigr\}, \\
& D_2 = 2 {\cal C}\Bigl\{[ k_{fT}^2 + (\hat h \cdot k_{fT})^2] [k_{iT} \cdot k_{fT} + 2 (\hat h \cdot k_{fT}) (\hat h \cdot k_{iT})]~{\rm F}(k_{iT},k_{fT}, \hat h) \Bigr\}, \\
& D_3 = 2 {\cal C}\Bigl\{[ k_{fT}^2 + 2(\hat h \cdot k_{fT})^2] [k_{iT} \cdot k_{fT} + (\hat h \cdot k_{fT}) (\hat h \cdot k_{iT})]~{\rm F}(k_{iT},k_{fT}, \hat h) \Bigr\}, \\
& D_4 = 2 {\cal C}\Bigl\{(\hat h \cdot k_{fT}) \bigl[(k_{iT} \cdot k_{fT}) (\hat h \cdot k_{fT}) + (\hat h \cdot k_{iT}) \bigl( k_{fT}^2 + 2(\hat h \cdot k_{fT})^2 \bigr) \bigr]~{\rm F}(k_{iT},k_{fT}, \hat h) \Bigr\}, \\
& D_5 = 2 {\cal C}\Bigl\{[ k_{fT}^2 + (\hat h \cdot k_{fT})^2] [k_{iT} \cdot k_{fT} + (\hat h \cdot k_{fT}) (\hat h \cdot k_{iT})]~{\rm F}(k_{iT},k_{fT}, \hat h) \Bigr\}.
\end{align}
\end{widetext}

\end{document}